END TO END DISTANCE ON CONTOUR LOOPS

OF RANDOM GAUSSIAN SURFACES

Moshe Schwartz
School of Physics and Astronomy
Tel Aviv University
Ramat Aviv, Tel Aviv 69978
Israel

A self consistent field theory that describes a part of a contour loop of a random Gaussian surface as a trajectory interacting with itself is constructed. The exponent $\nu$ characterizing the end to end distance is obtained by a Flory argument. The result is compared with different previous derivations and is found to agree with that of Kondev and Henley over most of the range of the roughening exponent of the random surface.

The study of surfaces, their static statistical properties, as well as growth and evolution dynamics has been attracting an ever increasing amount of interest over the last two decades. The main reason is that in addition to the many important problems of real surfaces arising in condensed matter physics,[1-6] this kind of problems is intimately related to other problems in physics like directed polymers,[2,7] string theory,[8] phase transitions in two dimensions[9] etc. In many of these studies the main interest lies in obtaining the distribution functional and the roughness exponent $\zeta$ characterizing these self affine surfaces. The purpose of the present letter is to study the statistical properties of contour lines of equal height. In particular I will be interested in the exponent $\nu$ characterizing the end to end distance $\Delta R$ of a part of such a contour of length S, $\Delta R \propto S^\nu$. The exponent $\nu$ will be discussed in the case of random surfaces that are governed by a Gaussian distribution. This is the simplest distribution one can think of, yet it may serve as a reasonable approximation of a real surface distribution, in particular, when one is interested mainly in large scale properties.[3]

The specific problem I consider in the following was discussed in the past using two different methods. The approximate multiscale analysis of Isichenko[10] yields $\nu = 7/(10-3\zeta)$, while very powerful arguments by Kondev and Henley[11] yield a different result, $\nu = 2/(3-\zeta)$.

The purpose of the present article is to construct a one dimensional field theory, that ascribes an energy to any given contour much in the way that it is done for polymers and self repelling polymers.[12] The main difference is that the "Hamiltonian" that ascribes an "energy" to any given contour $\vec{r}(t)$ is self consistent. Namely, its structure involves averages that have to be calculated with respect to the same Hamiltonian. The exponent $\nu$ obtained by this method agrees with that of Kondev and Henley[11] but only above $\zeta = 1/3$. Below that value the exponent I find from the self consistent Hamiltonian is just the end to end exponent of self avoiding walks but the use of the self consistent Hamiltonian in that regime is doubtful.

Consider a self affine surface $h = h(\vec{r})$, where $h$ is the height at the point $\vec{r} = (x,y)$ in the plane. The surface is assumed to be random and governed by a Gaussian distribution

$$P\{h\} \propto \exp[-\frac{k}{2}\int_o^{1/a} d^2q \ \ q^{2(1+\zeta)} h_q h_{-q}], \tag{1}$$

where $1/a$ is the high momentum cut-off and $h_q$ is the Fourier transform of $h(\vec{r})$. All the Fourier component of $h(\vec{r})$ with $q > 1/a$ are taken to be identically zero. This implies, with probability 1, that the surface is smooth on scales less or equal to $a$. In fact, if we extend $x$ and $y$ to the complex variables $z_1$ and $z_2$, respectively, $h(z_1, z_2)$ is an analytic function in both variables. Note also that since $h$ can be any physical scalar field, not necessarily a height, $a$ is the only length in the problem.

As a result of the smoothness of the surface, the contour lines are smooth on the same scale. Consequently, an arc length can be defined and as any other length it will be naturally measured in unite of $a$. My purpose is to study the statistical properties of contours of equal height, $\vec{r} = \vec{r}(s)$, in arc length intervals $o < s < S$, such that for $s_1, s_2$ in that interval $\vec{r}(s_1) \neq \vec{r}(s_2)$. In particular, I will be interested in the exponent $\nu$, that characterizes the end to end distance

$$< [\Delta \vec{r}(S)]^2 > \propto S^{2\nu}, \qquad (2)$$

where the average $< \; >$ is taken with respect to the distribution (1). The first goal is to obtain a statistical weight for a line starting at an arbitrary point and having the parametric representation $\Delta \vec{r} = \Delta \vec{r}(s)$. Note that the exponent $\nu$ is not expected to depend on a (universality) but non-universal quantities like prefactors will depend on $a$. For example, in eq. (2) above the constant of proportionality on the right hand side of the equation is a pure number times $a^{2(1-\nu)}$. Setting $a$ equal to zero renders $S^2$ and $< \Delta r^2 >$ infinite with a proportionality constant that is zero. Therefore, one can take the limit $a \to 0$ only for $\nu$ but not for $S$ or $< \Delta r (S)^2 >$.

An equal height contour can be obtained by following a "particle" that obeys the dynamical rule

$$\frac{d\vec{r}}{ds} = \hat{z} \times \frac{\nabla h}{|\nabla h|} \equiv \hat{z} \times \vec{u} \;, \qquad (3)$$

where $\hat{z}$ is a unit vector in the $z$ direction (perpendicular to the $xy$ plane) and $\vec{u}$ is a unit vector in the plane perpendicular to the line. The unit vector $\hat{z} \times \vec{u}$ is tangent to the equal height contour passing through $\vec{r}$. The fact that $\vec{u}$ is a unit vector implies that $s$ is indeed an arc length as required. The above "dynamical rule" takes the "particle" from any given starting point, along the equal height contour passing through that point with "speed" $V = 1$. The formulation of the problem becomes more tractable if we replace eq.(3) by

$$\dot{\vec{r}} \equiv \frac{d\vec{r}}{dt} = c\hat{z} \times \nabla h \;, \qquad (4)$$

where $c = [<|\nabla h|>]^{-1}$. Now, $</\dot{\vec{r}}/> = 1$ but we do not expect $/\dot{\vec{r}}/ = 1$ to hold for all $t$. Equation (4) above describes a particle moving with speed $V(\vec{r}) = |\nabla h(\vec{r})| / <|\nabla h|>$. I will be interested in the following in the exponent characterizing the dependence of $< \Delta r^2 >$ on the "elapsed time" T. How is this related to the dependence on S described earlier? It is straightforward to show, using the relation $S = \int_o^T V[\vec{r}(t)]dt$, that the average S for fixed T is T. For large T the fluctuation in S is expected to become vanishingly small compared to its average. This follows from the very plausible assumption, that the correlation $<(V[\vec{r}(t_1)]-1)(V[\vec{r}(t_2)-1)>$ vanishes for large $|t_2 - t_1|$, because a large "time" difference implies a large distance $|\vec{r}(t_2) - \vec{r}(t_1)|$. If this is the case, a choice of T defines actually a sharp S and the exponent characterizing the dependence of $<\Delta r^2>$ on T is the same characterizing its dependence on S. A proof that the above description is valid in our particular case is still lacking, however, and arguments suggesting a broad distribution of $S's$ for given $T$ can be presented. As long as a proof of either of the two possibilities does not exist it is important to note that for our purposes a much weaker result is needed. Indeed, even if the speed correlation tends to a constant, so that the fluctuation in S is of order T, S scales as T and the characterizing exponent is the same. (It is easy but somewhat lengthy to show that the fluctuation, $<\Delta S^2(T)>^{1/2}$ is bound from above by $cT$ T, where $c$ is a constant of order 1).

The probability of a given trajectory (I define one end point to be the origin)

$$\vec{r}(T) = \int_o^T \dot{\vec{r}}(t') \ , \tag{5}$$

is given by

$$P\{r(t)\} = < \prod_{t'=0}^{T} \delta(\dot{\vec{r}}(t') - c\hat{z} \times \nabla h(\vec{r}(t'))) \chi_T\{\vec{r}(t)\} >, \tag{6}$$

where $\chi_T\{\vec{r}(t)\}$ is a characteristic function that is zero if $\vec{r}(t_1) = \vec{r}(t_2)$ for some $t_1, t_2 < T$ and one otherwise. Since $\chi_T$ does not depend on the specific height configuration it can be brought out of the average and then can be described in terms of an Edwards interaction[12] so that

$$P\{\vec{r}(t)\} \propto < \prod_{t'=0}^{T} \delta(\dot{\vec{r}}(t') - c\hat{z} \times \nabla h(\vec{r}(t'))) > \times \exp\{-\frac{\lambda}{2}\int_o^T \int_o^T dt_1 dt_2 \delta[\vec{r}(t_1) - \vec{r}(t_2)]\} \tag{7}$$

Introducing a two dimensional vector field, $\vec{k}(t)$ and using the representation $\delta(x) = \frac{1}{2\pi}\int_{-\infty}^{\infty} e^{ikx} dx$, the probability $P\{\vec{r}(t)\}$ is found to be

$$P\{\vec{r}(t)\} \propto \int D\vec{k}(t) \ \exp[i\int_o^T dt \ \vec{k}(t) \cdot \vec{r}(t)] < \exp[-ic\int_o^T dt \vec{k}(t) \cdot (\hat{z} \times \nabla h)] > \times$$
$$\times \exp\{-\frac{\lambda}{2}\int_o^T \int dt_1 dt_2 \delta[\vec{r}(t_1) - \vec{r}(t_2)]\} \tag{8}$$

where $D\vec{k}(t)$ denotes functional integration over the vectorial function $\vec{k}(t)$ defined on the interval $0 \leq t \leq T$. Since $h$ is a Gaussian random field

$$< \exp[-ic\int_o^T dt \ \vec{k}(t) \cdot (\hat{z} \times \nabla h)] > =$$
$$= \exp[-\frac{1}{2}c^2 \int_o^T \int_o^T dt_1 dt_2 < \vec{k}(t_1) \cdot \hat{z} \times \nabla h[\vec{r}(t_1)] \vec{k}(t_2) \cdot \hat{z} \times \nabla h[\vec{r}(t_2)] >] \tag{9}$$

The needed average can be easily performed

$$<\vec{k}(t_1)\cdot\hat{z}\times\nabla h[\vec{r}(t_1)]\vec{k}(t_2)\cdot\hat{z}\times\nabla h[\nabla\vec{r}(t_2)]> = \vec{k}(t_1)\cdot(\hat{z}\times\nabla_1)\vec{k}(t_2)\cdot(\hat{z}\times\nabla_2)$$
$$<h[[\vec{r}(t_1)]-h[\vec{r}(t_2)]]^2> \equiv -\vec{k}(t_1)\cdot(\hat{z}\times\nabla)\vec{k}(t_2)\cdot(\hat{z}\times\nabla)w[\vec{r}(t_1)-\vec{r}(t_2)] \qquad , \qquad (10)$$

where $\nabla_1$ acts on functions of $\vec{r}(t_1)$ and $\nabla_2$ on functions of $\vec{r}(t_2)$.

Equations (8)-(10) define a one dimensional field theory in terms of the fields $\vec{k}(t)$ and $\vec{r}(t)$ governed by the Hamiltonian

$$H\{\vec{r},\vec{k}\} = \frac{c^2}{4}\int_o^T\int_o^T dt_1 dt_2 \Lambda[\vec{r}(t_2)-\vec{r}(t_1)]\vec{k}(t_1)\cdot\vec{k}(t_2)$$
$$+\frac{\lambda}{2}\int_o^T\int_o^T dt_1 dt_2 \delta[\vec{r}(t_1)-\vec{r}(t_2)] - i\int_o^T dt\ \vec{k}(t)\cdot\vec{r}(t) \quad , \qquad (11)$$

where $\Lambda = \nabla^2 w$.

In obtaining eq, (11) I have used $\hat{z}\times\vec{k}\to\vec{k}$ and rotational invariance.

The above Hamiltonian presents an exact field theoretic description of the problem, that should enable a systematic treatment. In the following I use, however, a simplification that will enable the $\vec{k}$ integration. The approximation I use is to replace $\Lambda[\vec{r}(t_2)-\vec{r}(t_1)]$ by its average over trajectories $\vec{r}(t)$. For large values of time separations, the distance, $\Delta r = |\vec{r}(t_2)-\vec{r}(t_1)|$, scales as the average over trajectories, so that inspite of a fluctuating prefactor of order 1, the replacement of $\Delta r$ by its average still yields a $\Lambda$ that has a correct scaling and this is what counts in the simple scaling (Flory) argument given in the following. (The fact that $\Lambda$ has the correct scaling at large time separations is a result of the fact that for large arguments $\Lambda$ behaves as a power law).
The remaining question whether indeed the long "time" dependence of $\Delta r$ is dominated only by long "time" separations in the Hamiltonian will be discussed later. At present I continue replacing $\Lambda[\vec{r}(t_2)-\vec{r}(t_1)]$ by its average for all time differences $|t_2-t_1|$.

The average $[\Lambda[[\vec{r}(t_1)-\vec{r}(t_2)]]]$ is a function of $|t_1-t_2|$ and going over to Fourier transform in $t$ facilitates the $\vec{k}$ integration. The result is a self consistent Hamiltonian for $\vec{r}(t), H\{\vec{r}(t)\}$, with parameters depending on averages, that have to be calculated with respect to the same Hamiltonian

$$H\{\vec{r}(t)\} = \frac{1}{2c^2}\int d\omega \frac{\omega^2}{\phi(\omega)}\ \vec{r}(\omega)\cdot\vec{r}(-\omega) + \frac{\lambda}{2}\int_o^T\int_o^T dt_1 dt_2 \delta[\vec{r}(t_1)-\vec{r}(t_2)], \qquad (12)$$

where $\vec{r}(\omega)$ is the Fourier transform of $\vec{r}(t)$ and $\phi(\omega)$ is the Fourier transform of $[\Lambda[\vec{r}(t_1)-\vec{r}(t_2)]]$. For small $\omega, \phi(\omega)\propto\omega^{-\mu}$ with

$$\mu = [1 - (2 - 2\zeta)\nu] \ . \tag{13}$$

Note that the above Hamiltonian is a generalization of the Hamiltonian of the self avoiding polymer that is characterized by $\mu = 0$.

I use now a Flory type argument to obtain a relation between $\Delta r$ and T. This is done by scaling: $\omega' = \omega T \quad \vec{r}' = \vec{r}/\Delta r$ and consequently $\vec{r}'(\omega') = \vec{r}(\omega)/(T\Delta r)$. As a result the first integral on the right hand side of eq. (12) scales as $(\Delta r^2)/T^{1+\mu}$ while the second integral scales as $T^2/(\Delta r)^2$. In order that the two terms be of the same order of magnitude we must have $(\Delta r)^4 \propto T^{3+\mu}$ or

$$\nu = (3 + \mu)/4 . \tag{14}$$

It is well known that although the use of a Flory type argument is not an exact treatment, it yields in many cases exponents that are extremely close to the exact ones and in some cases even the exact exponents.

Indeed, the exponents obtained from the Hamiltonian (12) for the case $\mu = 0$ (or $\mu < 0$ as will be seen later) is known to be exact.[13,14] Using eq. (13) one obtains

$$\nu = 2/(3 - \zeta) , \tag{15}$$

This is the result obtained by Kondev and Henley but only naively so. The reason is that, the small $\omega$ $\phi(\omega)$, is proportional to $\omega^{-\mu}$ with $\mu$ given by eq. (13), as long as that $\mu$ is positive. If there is no special reason to believe that indeed $\phi(\omega)$ can vanish for small $\omega$, then when eq. (13) yields a negative $\mu$ this means that $\phi(\omega)$ is actually a non zero constant at small $\omega$ (or that $\mu$ is actually zero). The end to end distance exponent of the self avoiding walk is thus recovered, $\nu = 3/4$.

The result of the above calculation is that the relation $\nu = 2/(3 - \zeta)$ holds only as long as $\zeta \geq 1/3$. For roughness exponents $0 \leq \zeta \leq 1/3$ $\nu = 3/4$ !.

The origin of the above small discrepancy is unclear yet . Its study as well as a more accurate treatment of the problem is postponed to future publications. At this stage I am in a position where I can only suggest the direction where the solution must be found. In the derivation of the self consistent Hamiltonian (eq. (12)), I have replaced $\Lambda[\vec{r}(t_2) - \vec{r}(t_1)]$ by its average. Clearly, this is valid only for large $|t_1 - t_2|$. The question is, how is the final result affected by the mistake done for small $|t_1 - t_2|$? Since the Hamiltonian $H(\vec{r}, \vec{k})$ (eq. 11) is quadratic in the k's, Gaussian integration yielding an exact Hamiltonian that depends only on $\vec{r}(t)$ is possible. The explicit form of the Hamiltonian can be written in terms of $\tilde{\Lambda}^{-1}$, the inverse of the kernel $\tilde{\Lambda}(t_1, t_2) = \Lambda[\vec{r}(t_2) - \vec{r}(t_1)]$. It proves useful to split the kernel into two parts: $\tilde{\Lambda}_1 = \tilde{\Lambda}[t_1, t_2] \times \theta[T_0 - |t_2 - t_1|]$ - and $\tilde{\Lambda}_2 = \tilde{\Lambda}[t_2, t_1] \times \theta[|t_2 - t_1| - T_o]$ where $T_o$ is a scale chosen so as to ensure that for $|t_2 - t_1| > T_o$, $\Lambda[\vec{r}(t_2) - \vec{r}(t_1)]$ may be replaced by its average.

The Fourier transform of the kernel

$$\hat{\Lambda}(\omega_1,\omega_2) = \frac{1}{T}\int_o^T\int_o^T dt_1 dt_2 \tilde{\Lambda}(t_2,t_1) e^{i(\omega_1 t_1 - \omega_2 t_2)} \qquad (16)$$

is thus the sum of two contributions, $\hat{\Lambda}_1(\omega_1,\omega_2)$, and $\hat{\Lambda}_2(\omega_1,\omega_2)$, where $\hat{\Lambda}_1(\omega_1,\omega_2)$ is bounded and $\hat{\Lambda}_2(\omega_2,\omega_1)$ is diagonal and has the form $\hat{\Lambda}_2(\omega_1,\omega_2) \propto \omega_1^{-\mu}\delta_{\omega_1\omega_2}$, for small $\omega_1$. Therefore, for $\mu > 0$ where $\hat{\Lambda}_2$ has eigenvalues that diverge as $\omega_1$ tends to zero, $\hat{\Lambda}_2$ provides an adequate description of the long time behaviour (as given by the self consistent Hamiltonian). Note, however, that $\hat{\Lambda}_2$ dominates over $\hat{\Lambda}_1$, provided $\mu > 0$. The use of the self consistent Hamiltonian is thus restricted to the regime $\varsigma > 1/3$. The final conclusion is therefore, that the Kondev-Henley relation is obtained here only for $\varsigma > 1/3$. It is unclear yet, what the present approach yields of $\varsigma < 1/3$.

The fact that the Kondev-Henley result is supposed to be exact at $\varsigma = 0$ [11,15] and the naive derivation of their result for all $0 \leq \varsigma \leq 1$ above (eq.(15)) pose intriguing problems that deserve further exploration.

**Acknowledgement.** This work was supported by an internal research fund of Tel Aviv University.